\documentstyle[12pt,epsfig]{article} 

\newcommand{\be}{\begin{equation}} 
\newcommand{\ee}{\end{equation}} 
\newcommand{\r}{\frac{r}{R}} 
\newcommand{\lp}{\left(} 
\newcommand{\rp}{\right)} 
\newcommand{\la}{\left\langle} 
\newcommand{\ra}{\right\rangle} 
\newcommand{\plateau}{{\it plateau}}
\newcommand{\ansatz}{{\it ansatz}}
\newcommand{\ie}{{\it i.e.}}

\title{Multiscale velocity correlation in turbulence: experiments, 
 numerical simulations, synthetic signals} 
\date{\today} 
\author{R. Benzi$^{1}$, L. Biferale$^{2,3}$, G. Ruiz-Chavarria$^{4}$,\\ 
S. Ciliberto$^{5}$ and F. Toschi$^{3,6}$ } 
\begin{document} 
%
%
%
%
\maketitle 

\centerline{$^{1}$ AIPA, Via Po 14, 00100 Roma, Italy.} 
\centerline{$^{2}$  Dipartimento di Fisica, Universit\`{a} di Tor Vergata} 
\centerline{Via della Ricerca Scientifica 1, I-00133 Roma, Italy.} 
\centerline{$^{3}$  INFM-Unit\'a di Tor Vergata.} 
\centerline{$^{4}$ Departamento de Fisica, Facultad de Ciencias, UNAM. 
04510  Mexico  D.F., Mexico} 
\centerline{$^{5}$ Laboratoire de Physique, URA 1325, Ecole Normale 
Sup\'erieure de Lyon.} 
\centerline{46 All\'ee d'Italie, 69364  Lyon, France} 
\centerline{$^{6}$ Dipartimento di Fisica, Universit\`a di Pisa, Piazza Torricelli 2, I-56126 Pisa, Italy.} 
\begin{abstract} 

Multiscale correlation functions in high Reynolds number experimental 
turbulence, numerical simulations  and synthetic signals are investigated.
Fusion Rules predictions as they arise from multiplicative, 
almost uncorrelated, random processes for the energy cascade are tested. 
Leading and sub-leading contribution, in  the inertial  range,
 can be explained as arising from a 
multiplicative random process for 
the energy transfer mechanisms. \\ Two different predictions
for  correlations involving dissipative observable  are also 
briefly discussed. 
\end{abstract} 

\newpage 
\section{Introduction} 
In stationary turbulent flows, a net flux of energy establishes in the 
inertial range, \ie\ from forced scales, $L$,  down to the 
dissipative scale $r_d$. 
Energy is transferred through a statistically scaling-invariant process, 
which is characterized by a strongly non-Gaussian (intermittent) 
activity. 
Understanding the statistical properties of intermittency is 
one of the  most  challenging open problem of three dimensional fully 
developed turbulence.\\ 

Intermittency in the inertial range  is usually analyzed by means of 
the statistical 
properties of  velocity differences, $\delta_r v(x) =v(x+r) -v(x)$. 
In particular, in the last twenty years \cite{frisch}, overwhelming 
experimental and theoretical works focused on structure functions: 
$S_p(r) =  \la (\delta_r v(x))^p \ra$. 
A wide agreement exists on the fact 
that structure functions show a scaling behavior in the limit of 
very high Reynolds numbers, \ie\ in presence of  a large 
separation between integral and dissipative scales, 
$L/r_d \rightarrow \infty$: 
\be 
S_p(r) \sim \lp {r \over L} \rp^{\zeta(p)} 
\label{structure} 
\ee 
The velocity fluctuations are anomalous in the sense that 
$\zeta(p)$ exponents do not follows the celebrated dimensional 
prediction 
made by Kolmogorov, $\zeta(p)=p/3$. In fact, 
 $\zeta(p)$ are observed  to be a nonlinear 
function of $p$, which is the most important signature of the 
intermittent transfer of fluctuations from large to small scales.\\ 
In order to better characterize the transfer mechanism, it is natural to 
look also at correlations among velocity fluctuations at different 
scales and at different times. The prototype of such a class of correlations is: 
\be 
C_{p,q}(r,R;\tau)=  \la (\delta_r v(x,t))^p (\delta_R v(x,t+\tau))^q \ra 
\label{timescalecorr} 
\ee 
where hereafter we will always assume the obvious notation: $ r < R$.\\ 
Unfortunately, the non-trivial time dependency of correlations 
such as (\ref{timescalecorr}) is completely hidden, in a Eulerian 
reference-frame, by the sweeping of small scales by large scales. 
The ``positive'' side of sweeping is connected to the Taylor hypothesis, 
 \ie\ to the possibility of identifying 
single point measurements at a time delay $\tau$  with 
single time measurements at separation scale $r \sim \tau  \bar{V}$ 
where  $\bar{V}$ is the large scale sweeping velocity.  The ``negative'' 
side of sweeping is connected to the fact that the inertial time-scales 
are always sub-dominant with respect to the sweeping time. 
This implies that in order to 
measure the temporal properties of the inertial 
range energy-transfer it is necessary to abandon the usual Eulerian 
reference-frame and to move in a Lagrangian or quasi-Lagrangian 
reference-frame where sweeping effects are absent 
 \cite{kraichan_lag,bel_lvov,bbct,pro1,lvov_pro}. Of course, 
from the experimental point of view, it is much harder to measure 
the velocity field in a Lagrangian frame than in the usual laboratory 
reference-frame. To our knowledge, the only results about 
multi-time velocity correlations are purely theoretical 
\cite{bbct,lvov_pro} or numerical \cite{bbct}. For this reason, 
in the following, we will limit to  an experimental and 
theoretical analysis of simultaneous -single time- 
multiscale correlation functions only. 

Multiscale correlations functions should play in turbulence the same 
r\^ole played by correlation functions in critical  statistical 
phenomena. 

\noindent 
Recently,  some theoretical work \cite{eyink,pro1,pro2} and an 
exploratory experimental investigation \cite{sreene} have been devoted 
to the behavior of single-time multiscale velocity correlations (MSVC): 
\begin{eqnarray} 
F_{p,q}(r,R) &\equiv&  \langle \lp v \lp x+r 
\rp-v \lp x\rp \rp^p \lp v\lp x+R\rp-v\lp x\rp\rp^q  \rangle  \nonumber \\ 
&\equiv& \langle (\delta_r v(x))^p(\delta_R v(x))^q \rangle  
\label{2s} 
\end{eqnarray} 
with $r_d <  r  <  R  < L$.  When 
the smallest among the two scales $r$ goes 
 beyond  the dissipative scales, $r_d$, new properties of the 
correlation functions (\ref{2s}) may arise due to the non trivial 
 physics of the dissipative cutoff \cite{frisch,physicad96}. From 
now on, we will mostly concentrate on correlation functions with both 
$r$ and $R$ 
in the inertial range, only in the last section 
we will address the important point concerned with the 
cross-over of the dissipative scale. Moreover, in order to simplify 
 our discussion, we will confine our analysis 
to the case of longitudinal velocity differences.\\ 
The main purpose of this paper is to review and to extend 
a recent experimental and theoretical analysis  of multiscale 
correlations (\ref{2s})  \cite{bbto}. In particular, 
we present a systematic analysis of multiscale 
correlation functions in different experimental set-up, we 
also perform a critical comparison with the same observable 
measured in synthetic turbulent signals defined in terms of 
purely multiplicative random processes. \\The comparison with the 
synthetic signals will allow us to conclude that 
multiscale correlation functions are in {\it quantitative} agreement, 
 with the prediction 
one obtains by using a pure uncorrelated multiplicative 
process for the energy cascade, as long as both separations $r,R$ are 
in the inertial range.\\ As for the case when one of the two separations 
is already in the dissipative range we will critically 
review the two most important different predictions one can obtain 
imposing the dissipative cut-off using either  multifractals 
 \cite{fv} or the GESS phenomenology \cite{physicad96}. Unfortunately, 
the actual state-of-the-art  experimental dissipative-scales data 
 does not allow 
to clearly distinguish among the two predictions \cite{adrienne,celani}. \\ 
The paper is organized as follows. In section 2 we briefly review the 
\ansatz\ that one simply obtains for MSVC (\ref{2s}) by using a multiplicative 
 random process 
for the inertial-range  energy cascade. In section 3 we discuss sub-leading 
corrections induced by 
geometrical constraints which necessarily affects any MSVC for  finite 
separation of scales. These geometrical constraints introduce sub-leading 
power laws behavior which may strongly interfere with the leading 
multiplicative predictions for finite separation of scales $r/R \sim O(1)$. 
In section 3 we also   present our experimental data-analysis 
and the comparison with the synthetic multifractal field. 
In section 4 we briefly address the problem of dissipative 
correlation functions. Conclusions follow in section 5. 

\section{Multiplicative random processes} 
Stochastic cascade processes  are simple and 
well known useful  tools to describe the leading phenomenology of the 
intermittent 
energy transfer in the inertial range. Both anomalous scaling 
exponents and viscous effects \cite{frisch,physicad96} can be 
reproduced by choosing a suitable 
random process for the multiplier, $W(r,R)$,  which connects 
velocity fluctuations  at two different scales, $R >  r$. \\ 

The main idea turns around the hypothesis that small scale statistics is 
fully determined by a cascade process conditioned  to  some 
large scale configuration: 
\be 
\delta_rv(x) = W(r,R) \cdot \delta_Rv(x) 
\label{linear} 
\ee 
where, requiring homogeneity along the cascade process, the 
random function $W$ should depend only on the ratio $r/R$. 
Structure functions are then described in terms of the $W$ process: 
$$S_p(r) = C_p  \langle  \left[ W\left( r/L \right)\right]^p  \rangle, $$ 
 where 
$C_p =  \langle  (\delta_{L}v(x))^p  \rangle $ 
if the stochastic multiplier may be 
 considered almost uncorrelated with the large-scale velocity field. 
Pure power laws arise in the high Reynolds regime: in this limit we must 
 have  $$ \la  \left[W\lp\r\rp\right]^p  \ra  \sim \lp\r\rp^{\zeta(p)}.$$ \\ 
In the same framework, it is straightforward 
 to give the leading prediction for 
the multiscale correlation functions (\ref{2s}): 
\begin{equation} 
F_{p,q}(r,R) \sim  \la 
 \left[ W\left( \rule[-2ex]{0cm}{4ex} \r \right) \right]^p 
\left[W \left(\frac{R}{L} \right)\right]^{p+q}  \ra, 
\label{frf} 
\end{equation} 
which becomes in the hypothesis of negligible correlations among 
multipliers: 
\begin{eqnarray} 
F_{p,q}(r,R) &=& C_{p,q}  \la  \left[W\left( \rule[-2ex]{0cm}{4ex} 
\r\right)\right]^p  \ra   
\la  \left[W\left(\frac{R}{L}\right)\right]^{p+q}  \ra  \sim\nonumber \\ 
&\sim& \frac{S_p(r)}{S_p(R)} \cdot S_{p+q}(R) 
\label{fritamar} 
\end{eqnarray} 
This expression was for the first time proposed in \cite{eyink} and 
later examined in more details in \cite{pro1} where it was named "fusion rules". In the same article the authors
proved  that the fusion rule prediction gives  the leading behavior of 
(\ref{2s}) when 
$r/R \rightarrow 0$ as long as some weak hypothesis of scaling 
invariance and of universality of scaling exponents  
in Navier-Stokes equations  hold. The name fusion-rules 
refers, probably, to the fact that thanks to the 
-supposed- uncorrelated nature of the 
cascade process  every multi-scale correlation can be written in terms 
of single scale correlations -\ie\ structure functions-.\\ 
Let us notice that, beside any rigorous claim, 
expression (\ref{fritamar}) is also the zero-{\it th} order prediction 
starting 
from any multiplicative uncorrelated random cascade satisfying 
$S_p(R)= \la  [W(\r)]^p \ra S_p(r)$.  Let 
us also stress that the fusion rules prediction  as stated in (\ref{fritamar}) 
  does not necessarily requires any scaling property 
of the underlying structure functions,  a fact which suggests that the 
validity of the statement  should be almost Reynolds independent. 
\\ 

In this paper we want to address three main questions: 
\begin{itemize} 
\item (i) whether the 
prediction (\ref{fritamar}) gives the correct leading behavior in  the 
limit of large scales separation: $r/R \sim 0$, 
\item (ii) if this is the case, what one can 
say about sub-leading behavior for  separation 
$r/R \sim O(1)$, 
\item (iii) what happens to those observable  for which the 
"multiplicative prediction" (\ref{fritamar}) is incorrect 
because of symmetry reasons. 
\end{itemize} 
The last item comes from the observation that for correlation like : 
\be 
F_{1,q}(r,R)=  \la \left(\delta_{r}v \right) \left(\delta_{R}v\right)^{q} \ra 
\label{fr1} 
\ee 
the multiplicative prediction gives: 
$$ F_{1,q}(r,R)= {{S_1(r)} \over {S_1(R)}} \cdot 
S_{1+q}(R).$$ 
 Such a prediction is 
wrong because, if homogeneity can be assumed, $S_1(r) \equiv 0$ for all scales 
$r$.  In this case prediction (\ref{fritamar}) 
does not represent the leading contribution. 

In the following  we propose a systematic investigation of (\ref{2s}) in 
high Reynolds number experiments \cite{exp1,exp2}, numerical 
simulation \cite{num2} and synthetic 
signals \cite{bbcpvv}. 
The main purpose consists in  probing whether 
multiscale correlation functions may show new dynamical 
properties (if any) which are not taken into account by the 
standard simple multiplicative models for the energy transfer. 

\section{Data analysis} 
In this section we present our main contributions by discussing 
the three items listed in the previous section and by presenting 
a detailed data-analysis in experiments at  different Reynolds numbers, 
in numerical simulations and in synthetic signals. 

Experimental data sets come mainly from two 
different laboratories. We have analyzed data  obtained in a wind 
 tunnel (Modane) 
with $Re_{\lambda}=2000$, the integral scale was 
$L\sim  20 \,m$ and the dissipative scale was $r_d = 0.31\, mm$. 
The second data set comes from a recirculating wind tunnel 
 (ENS de Lyon) with a working section 3 m long and 50x50 cm 
cross section. $Re_{\lambda}$  involved in experiments 
were $400$ (wake behind a cilinder) and $800$ (jet turbulence). 
Integral scales were $0.1m$ and $0.125m$ respectively 
whereas the dissipative scales were $0.15mm$ and $0.1mm$. 

Synthetic signals are built in terms of a 
Wavelet decomposition with coefficients defined by 
a pure uncorrelated random multiplicative process \cite{bbcpvv}. 
In the following,  the comparison between the synthetic field and  
  the experimental data will play a central  r\^ole in our discussion. 
 Therefore, in Appendix A we briefly recall how a multiaffine field may 
 be synthesized -and analyzed- in terms of a wavelet representation.\\ 
In appendix A we prove that such a   signal 
 shows the  fusion rules 
prediction (\ref{fritamar}) and therefore it will turn out to be an useful 
tool for testing how much deviations from (\ref{fritamar}), observed 
in experiments or numerical simulations, are due to important 
dynamical effects or only to unavoidable  geometrical corrections.

Let us proceed with a simple but basic observation. \\ 
For any notice that for any 1-dimensional string of number 
(such as the 
typical outcome of laboratory experiments in turbulence) the multiscale 
correlations (\ref{2s}) feel unavoidable strong geometrical 
constraints. In particular, for any MSVC, with two velocities at the same 
spatial point $v(x)$  and the 
two other velocities in a collinear geometry at spatial locations 
$v(x+r)$ and $v(x+R)$, like those analyzed in the 
following,  we will 
always write down  what we like to call  the "Ward-Identities" (WI): 
\begin{eqnarray} 
S_p(R-r) &\equiv&  \la \left[(v(x+R)-v(x))-(v(x+r)-v(x))\right]^p 
 \ra   \noindent \\ 
&=& 
\sum_{k=0,p}b(k,p)(-)^k F_{k,p-k}(r,R), 
\label{WI} 
\end{eqnarray} 
where $b(k,p)=p!/[k!(p-k)!]$.\\ 
For example, for $p=2$ we have 
\begin{eqnarray} 
2 F_{1,1}(r,R) &\equiv& S_2(r) + S_2(R) -S_2(R-r) \sim\nonumber \\ 
& \sim& \left[\left(\r\right)^{\zeta(2)}+ O\left(\r\right)\right] \cdot S_2(R) 
\label{f11} 
\end{eqnarray} 
where the latter expression has been obtained by expanding $S_2(R-r)$ 
in the limit $r/R \rightarrow 0$.\\ 
For $p=3$ we have 
$$S_3(R-r) = S_3(R) -S_3(r) +3 F_{2,1}(r,R) -3F_{1,2}(r,R)$$ 

The "Ward-Identities" will turn out to be useful for understanding 
sub-leading predictions 
to the multiplicative cascade process. One may argue that 
in geometrical set-up different from the one specified in (\ref{2s}) 
the same kind of 
constraint will appear with eventually different weights among 
different terms. 

The most important result one must extract from (\ref{WI}) is that 
the multiscale correlation functions, as stated in (\ref{2s}), may not be 
a perfect scaling functions even in the limit of very high Reynolds number. 
 Indeed, the WI tell us that MSVC with different 
order of velocity moments must be connected unavoidably one with the other: 
 which would be in contrast with the assumption that 
each MSVC should be determined by a single power law behavior. 

The main result presented in this work is that all multiscale 
correlations 
functions 
are well reproduced in their leading term, $\r \rightarrow 0$, 
by a simple uncorrelated random 
cascade (\ref{fritamar}) and that   their sub-leading contribution,  
$ \r \sim O(1)$, are fully captured 
by the geometrical constrained 
previously discussed, namely the "Ward-Identities". 
  
The recipe for calculating multiscale correlations is 
 the following: 
first, apply the multiplicative guess for the leading contribution and 
look 
for geometrical constraints in order to find out sub-leading terms.\\ 
Second, in all cases where the leading multiplicative contribution 
vanishes because of underlying symmetries, 
look directly for the geometrical constraints and find out what is 
the leading contribution 
applying the multiplicative random approximation to all, non-vanishing, 
terms  in the  WI. 

\subsection{Fusion rules: even moments} 

Let us  check the fusion rules prediction (\ref{fritamar}) for 
even moments  $p,q=2,4, \dots$. \\ 

In order to better highlight the scaling properties we will 
often use in the following, $\tilde{F}_{p,q}(r,R)$, the 
MSVC compensated with the fusion-rule prediction: 
\begin{equation} 
\tilde{F}_{p,q}(r,R) =\frac{F_{p,q}(r,R)\cdot S_{p}(R)}{S_p(r)\cdot S_{p+q}(R)} 
\label{comp} 
\end{equation} 

In order to compare experiments with different Reynolds numbers we may 
use as independent variable in our plot the quantity: 
$x(R) \equiv \frac{R-r}{L-r}$, 
where with $L$ we intend the integral scale of each different 
experiment. In this way, by fixing the small scale $r=5\eta$ and by changing 
$ r \le R \le L$ for each set of data we have a variation of 
$ 0 \le x(R) \le 1$. 
In Figure (1) we have checked the large scale dependency  by plotting the 
compensated MSVC  as a function of $x \equiv \frac{R-r}{L-r}$ at 
fixed, $r$, for $p=2,q=2$ and different Reynolds numbers (experimental 
and numerical). \\ 
The expression (\ref{fritamar}) predicts  the existence of a \plateau\ 
(independent of $R$) at all scales $ r \le R \le L$ 
where the leading multiplicative description is correct. \\

From Figure (1) one can see that experiments with low 
$Re$ numbers show a slightly poor \plateau. 
In particular the DNS (Direct Numerical Simulation) 
with  $Re_{\lambda}\sim 40$ does not show any \plateau. The absence 
of a \plateau\ is connected to the overwhelming geometrical effects 
present at such low Reynolds numbers (see below).\\ 
For this reason, in the following figures we will  show only 
 experimental data from Modane wind tunnel, 
which have the highest $Re$ number we can access. 

In Figure (2) 
one can see the compensated correlation functions for two different 
set of moments. In the limit of large separation $R \rightarrow L$ 
 at fixed $r$, 
we indeed see  a tendency toward a \plateau. On 
the other hand, there are clear deviations 
for $r/R \sim O(1)$. \\ 
The same behavior is seen in Figure (3) for 
the same compensated quantities fixing the large scale 
$R$ and   by changing the small scale $r$. \\ 
Such deviations show a very 
slow decay as a function of the scale separation. The decaying is 
so slow that a clear \plateau\ is seen only for the largest 
Reynolds number available. The question whether the 
observed finite-size 
corrections have an important physical origin or not is therefore 
of primary  importance. \\ 

In order to understand the physical meaning of the observed deviations 
to the fusion rules (\ref{fritamar}), 
we compare, in Figure (4), the experimental data against the equivalent 
quantities measured by using 
the synthetic signal. \\ 
We notice an almost perfect superposition 
of the two data sets, 
indicating that the deviations observed in real data 
can hardly be considered a "dynamical effect".\\ 

Using the WI plus the  multiplicative \ansatz\ for the leading behavior 
of all correlation functions 
for $p=4$ we quickly read that the leading contribution 
to $F_{2,2}$ is $ O(r^{\zeta(2)}) \cdot O(R^{\zeta(4)-\zeta(2)})$, while 
sub-leading terms scale as   $O(r^{\zeta(4)})$, and as 
$O(r^{\zeta(3)}) \cdot O(R^{\zeta(4)-\zeta(3)})$.\\ 
This superposition of power laws 
is  responsible for the slowly-decaying correlations in Figures (1-4).\\ 
In Table 1 we summarize the  leading and sub-leading contributions that may 
be inferred from the WI for the standard MSVC with $p=2,q=2$. Similar 
tables can  be constructed for any other even MSVC.\\ 
The result so far obtained, \ie\  that both the experimental data and the 
synthetic signal show the same quantitative behavior, is a strong indication 
that  multiscale correlation functions, at least for  even order 
of the moments -\ie\ in all cases where 
the signal is not affected by cancellation problems-, 
 are in good agreement 
with the random multiplicative model for the energy transfer. \\ 
An even stronger prove of this statement comes from the analysis 
of multiscale correlations in terms of the coefficients 
obtained by a Wavelet analysis of the experimental signal (see appendix A).\\ 
The Wavelet coefficient $\alpha_{j,k}$ may be seen as the representative 
of a velocity fluctuation at scale $r = 2^{-j}$  and centered 
in one of the $k=1,2,\cdots,2^{j}$ spatial point chosen 
equispaced in the original total length of the signal. \\ 
With this interpretation in mind, we may think at 
the Wavelet coefficients as the ideal observable which 
minimize the geometrical constraints and therefore as the ideal 
cases where one can test the idea that behind the multiplicative process 
there are only geometrical constraints. In other words,  in terms 
of the coefficients obtained by a Wavelet analysis of the 
experimental signal, the multiscale correlation function should 
show the fusion rules prediction for a range of scales much wider than 
for the velocity increments, \ie\ geometrical constraints, which 
introduce sub-leading power-laws decaying, should be minimized. \\ 
In Figure (5) we indeed prove that this is exactly what happens. In Figure (5) 
we show the equivalent of $F_{2,2}(r,R)$ built in terms 
of the Wavelet coefficients: 
\be 
F^{wav}_{2,2}(r=2^{-j},R=2^{-j'})= \la |\alpha_{j,k}|^2\,|\alpha_{j',k'}|^2 \ra 
\ee 
In the figure we plot, as in the previous figures, the compensated 
correlation, obtained from the Wavelet coefficients, 
 at fixed small scale  and at changing the large scale. 
In Figure (6) the same quantities are plotted at changing the small scale. 
As it is evident, the finite scale-separation  effects  visible in the standard 
MSVC have here  disappeared. \\ 
The \plateau\ is reached immediately after, 
say, one fragmentation step. 

\subsection{A case where fusion rules fail} 
For multiscale correlations where the direct 
application of the random-cascade 
prediction is useless -because of the translation symmetry-, 
like $F_{1,q}(r,R)$, we suppose that the main leading contribution is 
simply due to the geometrical constraints. In other words, we 
say that as soon as the main leading effect induced by the presence 
of a multiplicative  random energy transfer is depleted because 
of symmetry reasons, the sub-leading  contributions  induced by the geometry 
becomes the leading contributions. \\In order to give a prediction 
for such class of MSVC we therefore use 
the WI applying the  multiplicative prediction  to all terms, except 
the $F_{1,q}$.\\ One obtains the  expansion: 
\begin{eqnarray} 
F_{1,q}(r,R) &\sim& \left[ O\left( \r \right)^{\zeta(2)} + 
O\left(\r \right)^{\zeta(3)} 
+O\left(\r\right)^{\zeta(4)} + \nonumber \right. \\ 
 &\cdots& \left.+ \; O\left(\r\right)^{\zeta(q+1)} \right] \cdot S_{q+1}(R), 
\label{f1q} 
\end{eqnarray} 
which coincides when $q=1$ with the exact result (\ref{f11}) using 
$\zeta(3)=1$. 

In Figure (7) we show the experimentally measured $F_{1,2}$ and the fit that 
we obtain by keeping only the first two terms of the 
expansion in (\ref{f1q}). The fit has been performed by  imposing the 
value for the scaling exponents $\zeta(2), \, \zeta(3)$ measured 
on the structure functions, \ie\ only the coefficients 
in front of the power laws have been fitted. 
As one can notice, the fit 
works perfectly in the inertial range. 
Let us remark that 
the correlation changes sign in the middle 
of the inertial range, which 
is a clear indication that a single power-law 
fit (neglecting sub-leading terms) would completely miss the correct 
behavior. 

Next we consider the WI for $p=3$. Due to the fact that $S_3(r) \sim r$ 
in the inertial range, 
one can easily show that the WI 
enforces $F_{12} \sim F_{21}$. Therefore we can safely state that 
also correlation functions of the form $F_{p,1}$ 
feel non trivial dependency from the large scale $R$, 
proving that the  prediction given in \cite{pro2} 
using isotropy arguments is wrong. 

\subsection{Fusion rules: odd moments} 

For the most general MSVC involving odd moments of velocity 
increments, $F_{p,q}(r,R)$ with $p,q=3,5,7, \dots$.  the situation is 
 slightly more confused. 
The problem stems from the fact that the  fusion rules contribution to this 
 correlation feels indeed the skewed part of the process which is 
order of magnitudes less important than the even part. For example,  the 
 multiplicative contribution to the correlation $F_{3,1}(r,R)$ 
would be $S_3(r)/S_3(R) \cdot S_4(R)$ which is different form zero 
only due to the fact that the  process for the longitudinal 
velocity correlation is skewed.  \\ 
The weakness of the signal from the 
multiplicative contributions makes these class of correlation functions 
very hard to analyze from the point of view of scaling. Here, the geometrical 
 constraints may well be more important, in a large 
range of scale separation, than the fusion rules prediction. For example 
in Figure (8) we plot the standard MSVC for $p=3,q=1$ and the same 
correlations but with moduli of velocity increments, such as to 
get rid, in the second case, of cancellation effects. As it is evident, the 
two correlation have a very different amplitude as soon as 
the scale separation becomes important and it is hard 
to say whether the MSVC without moduli follow the fusion rules 
prediction for large scale separation or not. On the other hand, 
the correlation with absolute values does follow the 
multiplicative prediction reaching a \plateau\ after the usual 
finite size transient as the ordinary even-MSVC. \\ 

A high statistics and high Reynolds number investigation of such a 
class of correlation may well be of some interest in order to elucidate 
whether the odd part of velocity increments follows the same 
physics of the even part or not. 

\section{Dissipative Physics} 

In this section we discuss the application of fusion rules in the 
dissipative range. We will be mainly interested in the following two
quantities:
\be
A_n(R) = \la \Delta v(x) \cdot \delta v_R(x)^n \ra 
\label{A}
\ee
\be 
B_{p,n}(R) = \la T(x)^p \cdot \delta v_R(x)^n \ra
\label{B}
\ee
where $\Delta v(x)$ is the laplacian computed at the point $x$,
$\delta v_R (x) = v(x+R) - v(x)$ and $T(x)$ is the velocity gradient
computed at $x$.
In order to simplify the discussion we restrict to the one dimensional case,
namely the Laplacian and the gradient  are 
computed in one dimension and velocity differences are 
longitudinal. Moreover we restrict our analysis to
the cases of $n$ odd and $n+p$ even.
Our findings will anyway be valid in the most general case.
The scaling properties of $A_n$ and $B_{p,n}$ have been investigated in
\cite{pro1}, \cite{adrienne} and \cite{bbto}.

We start by considering the scaling properties of $A_n$.
By its definition we have:

\begin{eqnarray}
\nonumber  A_n(R) &=& \lim_{r \rightarrow 0} \la \lp \frac{v(x+r)+v(x-r)-2v(x)}{r^2}
\rp  \cdot \left( v(x+R)-v(x) \right)^n \ra = \\
&=& \lim_{r \rightarrow 0}  r^{-2}\left( F_{1,n}(r,R)+F_{1,n}(-r,R)\right)
\label{def}
\end{eqnarray}
In order to understand how equation (\ref{def}) works, we compute the easiest
quantities, \ie\  $A_1$ and $A_3$. By using equation (\ref{f11}), we obtain:
\begin{eqnarray}
A_1(R) &=& \lim_{r \rightarrow 0}   
\frac{r^{-2}}{2}(2S_2(R) + 2S_2(r) - S_2(R-r) -S_2(R+r)) = \\
&=&\lim_{r \rightarrow 0}  r^{-2}
\left(S_2(r)-\frac{1}{2}r^2\frac{d^2S_2(R)}{dR^2} + 0\lp r^3\rp\right) = \\
&=& \la (\partial_x v)^2 \ra -\frac{1}{2}\frac{d^2S_2(R)}{dR^2}
\label{A1}
\end{eqnarray}
In equation (\ref{A1}), we have used the relation:
\be
\la \left(\partial_xv\right)^2 \ra = \lim_{r \rightarrow 0} \frac{S_2(r)}{r^2}
\ee
The computation of $A_3$ is similar and we find, using (\ref{WI}):
\begin{eqnarray}
\nonumber A_3(R) &=&  \lim_{r \rightarrow 0}  r^{-2}
\nonumber \frac{1}{4}(2S_4(R)+2S_4(r) + 12F_{2,2}(r,R) - 4 F_{3,1}(r,R) +\\
\nonumber &&-4F_{3,1}(-r,R) -S_4(R-r)-S_4(R+r)) = \\
\nonumber &=&\lim_{r \rightarrow 0}  r^{-2}(3F_{2,2}(r,R)-\frac{1}{4}r^2\frac{d^2S_4(R)}{dR^2}+ O\lp r^3\rp) =  \\
&=&3B_{2,2}(R) -\frac{1}{4}r^2\frac{d^2S_4(R)}{dR^2} 
\label{A3}
\end{eqnarray}
In equation (\ref{A3}) we used the definition of $B_{2,2}$, namely
\be
B_{2,2}(R) =  \lim_{r \rightarrow 0} r^{-2} F_{2,2}(r,R)
\ee
At this point it is quite easy to find the most general expression for
$A_n$, which is
\be
A_n(R) = nB_{2,n-1}(R) -\frac{1}{n+1}\frac{d^2S_{n+1}(R)}{dR^2}
\label{An}
\ee
Equation (\ref{An}) is an exact results which is independent on any physical
assumption on the fusion rules. The last term on the r.h.s. of (\ref{An})
becomes small for $R$ in the inertial range. On the other hand,
for small value of $R$, \ie\ for $R \rightarrow 0$, the last term of the
r.h.s. of (\ref{An}) cannot be neglected.
In particular, for $R \rightarrow 0$, 
an explicit computation,
either using (\ref{An}) or (\ref{A}), gives -after cancellations of 
leading terms in the RHS of (\ref{An})-:
\be
A_n(R) \simeq O\lp R^{n+1}\rp
\label{vincolo}
\ee
In order to complete our computation for $A_n$, we need an estimate
for $B_{2,n}$. There are in principle two ways to compute $B_{p,n}$:
the first one using the multiscaling approach 
\cite{fv}, the second one using the GESS theory discussed
in \cite{physicad96}. 

We first analyze the case of multiscaling. In this case,  
one can use the approach of multiplicative processes with multiscaling 
viscous cutoff \cite{fv}. Namely, for the correlation 
$B_{2,n}(R)=\la\lp\partial_xv\rp^{2} \lp\delta_R v(x)\rp^n\ra$ one 
obtains: 
\begin{equation} 
B_{2,n}(R) \sim  \la  \lp \delta_R v(x) \rp^n  
\lp \frac{\delta_{r_d}v}{r_d}\rp^2 \ra 
\end{equation} 
where $r_d$ is the dissipative scale. 
In the multifractal interpretation we assume: 
$\delta_{r_d}v = (r_d/R)^h \cdot \delta_R v$ 
with probability $P_h(r_d,R)= (r_d/R)^{3-D(h)}$. 
 Following \cite{fv} we have: 
\begin{equation} 
\delta_{r_d}v \cdot r_d \sim 
\left(\frac{r_d}{R} \right)^h \delta_R v \cdot r_d \sim \nu. 
\end{equation} 
Inserting the last expression in the definition of $B_{2,n}(R)$, we 
finally have: 
\begin{equation} 
B_{2,n}(R) \sim \int d\mu(h) {{\lp\delta_R v \rp^{n+2}} \over {R^{2}}} 
 \ 
\cdot \left( \frac{\nu}{R \cdot \delta_R v} 
\right)^{\frac{2 \left( h-1\right)+3-D \left(h \right)}{1+h}} 
\sim  \frac{S_{n+3}(R)}{\nu \cdot S_3(R)} 
\label{primocaso} 
\end{equation} 
where we have used the fact that the multifractal 
process is such that \\ 
 $\nu \la (\partial_x v)^2 
 \ra \rightarrow O(1)$ 
in the limit $\nu \rightarrow 0$. Expression 
(\ref{primocaso}) coincides with 
 the prediction given in \cite{pro2}. 
 The above 
computation are easily 
generalized for any  
$\la  (\partial_xv)^{p} (\delta_R v(x))^q  \ra$.
By using (\ref{primocaso}) and (\ref{An}) we finally obtain:
\be
A_n(R) = n C_n \frac{S_{n+2}(R)}{ \nu S_3(R)} - \frac{1}{n+1}\frac{d^2S_{n+1}(R)}{dR^2}
\label{ris1}
\ee
Let us note that, for $R \rightarrow 0$,
equation (\ref{ris1}) predicts that $A_n(R) \sim O\lp R^{n-1}\rp$ which violates equation (\ref{vincolo}).

We now compute $A_n(R)$ by using the GESS approach discussed in
\cite{physicad96}.
In this case the computation of $B_{2,n}$ can be
easily done by noting that, within the GESS approach, the fluctuations
of the dissipation scale
are confined in the range where $\delta_R v \sim R$. This implies that,
for what concerns the scaling properties of $B_{2,n}(R)$, the effect of
a fluctuating dissipation scale can be disregarded. Following
\cite{physicad96}, after a long but straightforward computation, we
obtain:
\be
B_{2,n}\lp R\rp = D_n \frac{\la T^2 \ra S_{n+2}\lp R\rp}{S_2\lp R\rp}
\label{secondocaso}
\ee
where $T$ is the velocity gradient and $D_n$ is a constant.
Equation (\ref{secondocaso}) can be easily understood by noting that,
within the GESS approach,
$F_{2,n}(r,R) \sim \frac{S_2(r)\cdot S_{n+2}(R)}{S_2(R)}$ for any values
of $r$ and $R$, \ie\ also in the limit $r \rightarrow 0$.
Using (\ref{secondocaso}) we finally obtain:
\be
A_n(R) = n D_n  \frac{\la T^2 \ra S_{n+1}(R)}{S_2(R)} - \frac{1}{n+1}\frac{d^2S_{n+1}(R)}{dR^2}
\label{ris2}
\ee
For $R \rightarrow 0$, using the estimate $S_n\lp R\rp \sim \la T^n \ra
R^n + O\lp R^{n+2}\rp$, and the fact that $D_n = 1$ for $R = 0$, we can easily
show that equation (\ref{ris2}) satisfies the constrain (\ref{vincolo}).

From an experimental point of view, it is extremely difficult
to distinguish between the two predictions
(\ref{ris1}) or (\ref{ris2}). We note that the experimental and numerical 
analysis 
discussed in \cite{adrienne,celani}, has been done neglecting the second term
on the r.h.s. of (\ref{An}). Also, the experimental
analysis performed in \cite{physicad96} seems to indicate that
multiscaling effects are not observed in real turbulence.
At any rate, no definitive conclusions can be drawn from existing
experimental data.

\section{Conclusions} 
Let us summarize what is the framework we have found until now.\\ 
Whenever the simple scaling \ansatz\ based on the 
uncorrelated multiplicative process is not prevented by 
symmetry arguments, the multi-scale correlations are in good 
asymptotic agreement with the fusion rules prediction even if 
strong corrections due to sub-leading terms are 
seen for small-scale separation $r/R \sim O(1)$. 
Sub leading terms are strongly connected to the WI  previously 
discussed, \ie\ to geometrical constraints. 
In the other cases (\ie\ $F_{1,q}(r,R)$) 
the geometry fully determines both leading and  sub-leading scaling.\\ 
All these findings, led us to the conclusions that multiscale 
correlations 
functions measured in turbulence are fully consistent 
with a multiplicative, almost uncorrelated, random process.\\ 
Nevertheless, the strong and slowly-decaying sub-leading corrections 
to  the naive multiplicative 
fusion rules predictions are particularly annoying for any attempts  to 
attack 
analytically  the equation of  motion for structure functions; 
in that case, multiscale 
correlations at almost coinciding scales 
are certainly the dominant contributions 
in the non-linear part of the equations \cite{pro2}. Indeed, as 
shown in an  analytical calculation 
for a dynamical toy model of random passive-scalar advection \cite{bbw}, 
fusion rules are violated at small scale-separation and 
the violations are relevant for correctly evaluating the exact behavior 
of structure functions at all scales. 

Finally, let us remark that the standard multiplicative 
process may not be the end of the story, 
\ie\ the dynamics  may be more complex 
than what here summarized.\\ 
 For example, one cannot exclude that 
also  sub-leading (with respect to the 
multiplicative \ansatz) dynamical processes are acting in the energy 
transfer from large to small scales. These dynamical corrections 
 must be either negligible with respect to the geometrical 
constraints or, at best, of the same order. Also, as shown in this 
paper, the odd correlation functions are  not 
 jet under control: higher Reynolds number experiments, with 
higher statistics, are needed. 

The question connected to the transfer properties of 
quantities "orthogonal" to the energy may reveal different physical 
mechanisms \cite{bpt}. What happens for all those multiscale 
correlation functions which feel a non-trivial helicity dependency 
for non-parity invariant flows is in this framework  an open question. 

For what concerns fusion rules involving
velocity gradients or laplacian and
velocity differences, we observe that there are controversial
arguments leading to different predictions. It is difficult to
distinguish which predictions is really observed in real turbulence,
because experimental data at large Reynolds number do no resolve
the far dissipative range with enough accuracy.

Other possible candidates to investigate the previous problems are 
shell models for turbulence, where 
geometrical constraints do not affect the energy cascade mechanism. 

We acknowledge useful discussions with A.L. Fairhall, V. L'vov and I. 
Procaccia. M. Pasqui is kindly acknowledged for his help 
in the analysis of the  synthetic signal. 
We are indebted to 
 Y. Gagne for having allowed us 
the access to the experimental data. 
L.B. and F.T have been supported by INFM (PRA TURBO). G. R. C. and S. C. 
acknowledge support by ECOS comitee and CONACYT under project M96-E03.

\section{Appendix A} 
We  build  up  a 1-dimensional synthetic signal 
according to a random multiplicative process defined in a dyadic 
 hierarchical structure as originally introduced in \cite{bbcpvv} (for 
a review and references see also \cite{sreene_sint}).\\ 
Let us consider a wavelet decomposition of the  function $ \phi (x)$: 
\begin{equation} 
\phi(x)=\sum_{j,k=0}^{\infty} \alpha_{j,k} \psi_{j,k}(x) 
\label{6.26} 
\end{equation} 
where $\psi_{j,k}(x)= 2^{j/2} \psi(2^jx -k)$ and $\psi(x)$ is any wavelet 
with zero mean. The above decomposition defines the signal 
as a dyadic superposition of basic fluctuations with different 
characteristic widths (controlled by the index $j$) and centered in 
different spatial points (controlled by the index $k$). 
For functions defined on $N=2^n$ points in the interval $[0,1]$ 
the sums in (\ref{6.26}) are restricted from zero to $n-1$ for 
the index $j$ and from zero to $2^j-1$ for $k$. 

In \cite{bbcpvv} it has been shown that the statistical behavior of 
signal increments: 
$$<|\delta \phi(r)|^p> = 
<|\phi(x+r)- \phi(x)|^p> \sim r^{\zeta(p)}$$ 
is controlled by the coefficients $\alpha_{j,k}$. By defining the 
 $\alpha$ coefficients in terms of 
  a multiplicative random process on the dyadic tree 
  it is possible to give an explicit expression for the scaling 
  exponents $\zeta(p)$. For example,  it is possible to recover 
  the standard anomalous scaling by defining the $\alpha$'s tree in term 
  of the realizations of a random variable $\eta$ with a probability 
   distribution 
  $P(\eta)$: 
$$ \alpha_{0,0}$$ 
  $$ 
\alpha_{1,0}  =   \, \eta_{1,0} \, \alpha_{0,0}; \,\,\, 
\alpha_{1,1}  =  \, \eta_{1,1} \, \alpha_{0,0}; \nonumber $$ 
\begin{equation} 
\alpha_{2,0} = \,\eta_{2,0} \,\alpha_{1,0}; \,\, 
\alpha_{2,1} = \,\eta_{2,1} \,\alpha_{1,0}; \,\, 
\alpha_{2,2} = \,\eta_{2,2} \,\alpha_{1,1}; \,\, 
\alpha_{2,3}=  \,\eta_{2,3} \,\alpha_{1,1}, 
 \label{eq:alb} 
\end{equation} 
and so on. Let us note that in the previous multiplicative 
process different scales are characterized by 
different values of the index $j$, \ie\ $r_j =2^{-j}$. 
If the $\eta_{j,k}$ are indipendent identically distributed random variable
 it is straightforward to realize that $\alpha_{j,k}$ are random variables 
with moments given by: 
\begin{equation} 
<|\alpha_{j,k}|^p> = r_j^{- \log_2(\overline{\eta^p})} = r_j^{\zeta(p)} 
\label{eq:scal_alfa} 
\end{equation} 
where   the ``mother eddy' $\alpha_{0,0}$ 
 has been chosen equal to one. In (\ref{eq:scal_alfa}) 
 with $\overline{\cdots}$ we intend averaging over the $P(\eta)$ 
distribution. 
In \cite{bbcpvv} it has been shown that also the signal $\phi(x)$ 
 has the same  anomalous  scaling of (\ref{eq:scal_alfa}).\\ 
The same arguments used in order to prove 
that the field $\phi(x)$  has an anomalous scaling can be invoked to 
  show also that the  fusion-rules prediction (\ref{fritamar}) are 
satisfied -at least for large scale separation-. \\ 
On the other hand, it is a trivial matter to realize that the above signal 
will show {\it exactly}, and {\it for any} separation of scale, the 
fusion-rules prediction if expressed for the wavelet coefficients 
$\alpha_{j,k}$. For example, let us consider two wavelet coefficients at 
different scales $r_j < r_{j'}$ and let us chose the $k$-s indices such that 
the two coefficients refer to two spatially overlapping wavelets, then 
it is trivial to realize that, due to the multiplicative nature of the 
wavelet coefficients, we have: 
\be 
\la \left|\alpha_{j,k}\right|^p\,\left|\alpha_{j',k'}\right|^q \ra 
\equiv \left(\frac{r_j}{r_{j'}}\right)^{\zeta(p)} r_{j'}^{\zeta(p+q)} \ 
\label{fusrule_wave} 
\ee 
which shows that the fusion rules prediction is satisfied 
exactly for any separation of scales as long as the two fluctuations 
are chosen with overlapping distances. In the case the two distances 
are not overlapping, deviations from the fusion rules prediction are 
certainly seen in the synthetic field due to the dyadic -ultrametric- nature 
of the underlying structure. The question whether such deviations may be 
seen also in the experimental data is an interesting point which 
is outside the scope of this paper (see for similar problems 
\cite{bbt_ultra,mene_ultra}).

\newpage 

\centerline{FIGURES and TABLE CAPTIONS} 

\noindent 
{\bf TABLE 1}:\\ 
Leading (first column) and sub-leading (second column) contribution 
to the different multi-scale velocity correlations entering 
in the WI written for $p=4$. Notice that all the leading behaviors 
have been obtained by using the multiplicative \ansatz\ (when applicable). 
 The sub-leading behaviors are consistent with the constraints imposed 
by the WI.

\vskip 0.5 truecm
\noindent 
{\bf FIGURE 1}:\\ 
Compensated MSVC 
$ \tilde{F}_{2,2}(r,R)$ 
at fixed $r$ and changing $x(R)= \frac{R-r}{L-r}$ 
 for different experiments and numerical simulation: 
($\times$) Direct Numerical 
Simulation ($Re_{\lambda}= 40$), ($+$) Jet ($Re_{\lambda}= 800$), 
 ($ \ast $) Modane ($Re_{\lambda}= 2000$), ($ \Box $) Wake ($Re_{\lambda}= 400$). 

\vskip 0.5 truecm
\noindent 
{\bf FIGURE 2}:\\ 
Compensated MSVC 
$\tilde{F}_{p,q}(r,R)$ 
at fixed $r$ and changing the large scale $R$ for $p=2, q=2$ ($+$) 
and $p=2, q=4$ ($\times$).

\vskip 0.5 truecm
\noindent 
{\bf FIGURE 3}:\\ 
Experimental compensated MSVC 
$ F_{p,q}(r,R)/S_{p+q}(R) \cdot S_p(r)$ 
at fixed $R$ and changing the  small scale $1/r$ 
 for $p=2, q=2$ ($+$) and $p=2, q=4$ ($\times$). 

\vskip 0.5 truecm
\noindent 
{\bf FIGURE 4}:\\ 
Comparison between experimental   and synthetic 
compensated MSVC, $\tilde{F}_{p,q}(r,R)$ 
at fixed $r$ and changing the  large  scale $R$ 
for $p=2, q=2$: ($+$) synthetic and ($\times$) experimental. 
For $p=2, q=4$: ($\ast$) synthetic and ($\Box$) experimental. 

\vskip 0.5 truecm
\noindent 
{\bf  FIGURE 5}:\\ 
Comparison between real space ($+$) and wavelet analysis ($\times$) 
 of the 
 experimental data set from Modane. Compensated $\tilde{F}_{2,2}$ 
 is shown for fixed $r$ at varying $R$. 

\vskip 0.5 truecm
\noindent 
{\bf  FIGURE 6}:\\ 
Comparison between real space ($+$) and wavelet analysis ($\times$) 
of the  experimental data set from Modane. 
Compensated $\tilde{F}_{2,2}$ is shown for fixed $R$ at varying $r$. 

\vskip 0.5 truecm
\noindent 
{\bf  FIGURE 7}:\\ 
Experimental $F_{1,2}(r,R)$ at fixed 
$r= 16\,r_d$ and at varying $R$. 
The integral scale $L \sim 1 \times 10^4 \, r_d$. 
Let us remark that the observed 
change of sign in the correlation implies 
the presence of at least two power laws. 
The continuous  line 
is the fit in the 
region $r  <  R  < L$ obtained by 
using only the first two terms in (\ref{f1q}). 

\vskip 0.5 truecm
\noindent 
{\bf  FIGURE 8}:\\ 
Comparison between compensated $\tilde{F}_{3,1}$ odd MSVC with absolute 
values ($+$)  and without ($\times$). 
 Data  are shown for fixed $r$ at varying $R$. 
It is evident that the odd MSVC with absolute values 
has the same behavior of even MSVC, while the one without 
 absolute value does not follow the same behavior. 

%
%
%
%
%
%
%
%
%
%
%
%
%
%
%

\end{document}